\documentclass[aps,prd,twocolumn,nofootinbib,superscriptaddress]{revtex4}
\usepackage[hypertex]{hyperref}
\usepackage{graphicx}
\def\lesssim{\mathrel{\mathpalette\vereq<}}
\def\gtrsim{\mathrel{\mathpalette\vereq>}}

\begin{document}

\pagestyle{plain}

\title{The Minimal Supersymmetric Fat Higgs Model}

\author{Roni Harnik}
\affiliation{Theoretical Physics Group, 
Ernest Orlando Lawrence Berkeley National Laboratory, 
University of California, Berkeley, CA 94720}
\affiliation{Department of Physics, University of California,
Berkeley, CA 94720}

\author{Graham D. Kribs}
\affiliation{\mbox{Institute for Advanced Study, Princeton, NJ 08540} 
\\[2mm]
\mbox{\textnormal{\texttt{
roni@socrates.berkeley.edu,
kribs@ias.edu,
dtlarson@socrates.berkeley.edu,
murayama@ias.edu
}}}}

\author{Daniel T. Larson}
\affiliation{Theoretical Physics Group, 
Ernest Orlando Lawrence Berkeley National Laboratory,
University of California, Berkeley, CA 94720}
\affiliation{Department of Physics, University of California,
Berkeley, CA 94720}

\author{Hitoshi Murayama}
\thanks{On leave of absence from Department of Physics, 
University of California, Berkeley, CA 94720.}
\affiliation{\mbox{Institute for Advanced Study, Princeton, NJ 08540} 
\\[2mm]
\mbox{\textnormal{\texttt{
roni@socrates.berkeley.edu,
kribs@ias.edu,
dtlarson@socrates.berkeley.edu,
murayama@ias.edu
}}}}

\date{\today}

\begin{abstract}
We present a calculable supersymmetric theory of a composite ``fat''
Higgs boson. Electroweak symmetry is broken dynamically through a new
gauge interaction that becomes strong at an intermediate scale. The
Higgs mass can easily be 200--450~GeV along with the superpartner
masses, solving the supersymmetric little hierarchy problem.  We
explicitly verify that the model is consistent with precision
electroweak data without fine-tuning.  Gauge coupling unification can
be maintained despite the inherently strong dynamics involved in
electroweak symmetry breaking.  Supersymmetrizing the Standard Model
therefore does not imply a light Higgs mass, contrary to the lore in
the literature.  The Higgs sector of the minimal Fat Higgs model has a
mass spectrum that is distinctly different from the Minimal
Supersymmetric Standard Model.
\end{abstract} \pacs{Who cares?} \maketitle

\section{Introduction}
\label{sec:intro}

The mechanism and dynamics of electroweak symmetry
breaking~\cite{Salam:rm} is one of the greatest mysteries in particle
physics. The Standard Model (SM) provides an extremely successful
parameterization of electroweak symmetry breaking through the
introduction of a single Higgs doublet with a negative mass squared.
Furthermore, fermion masses, mixings and CP violation are accommodated
in a way that is consistent with experimental observations and
constraints from flavor-changing neutral current processes.
Unfortunately, the negative mass-squared is just an input to the
theory without a deeper understanding.  Moreover, symmetry breaking by
relevant operators is fraught with an extreme ultraviolet (UV)
sensitivity: the Higgs mass squared is quadratically sensitive to new
physics. Thus the origin of electroweak symmetry breaking in the SM is
impossible to understand without a complete UV theory.

It is well known that supersymmetry removes the extreme quadratic
sensitivity to the UV and can explain electroweak symmetry breaking
with clear predictions for experiments. The Minimal Supersymmetric
Standard Model (MSSM), in particular, predicts a light Higgs boson,
$m_h \lesssim 130$~GeV, for top-squark masses less than about 1 TeV\@.
Superpartners, especially charginos and stops, are also expected to be
light because their masses feed into the Higgs mass parameters via the
renormalization group \cite{Barbieri:1987fn}. The experimental lower
bounds on the mass of the lightest Higgs boson and the masses of the
superpartners are increasingly becoming a quantifiable concern. Even
though the MSSM is far from being excluded, it already relies on
fine-tuning at the level of a few percent: we call this the
``supersymmetric little hierarchy problem''.

The simplest way around this problem is to raise the Higgs mass. In
the MSSM the Higgs mass could be raised by increasing the top squark
masses, but only with an unnaturally drastic increase in fine-tuning.
A less fine-tuned approach is to invoke physics beyond the MSSM that
provides additional contributions to the quartic coupling.  The
simplest idea of this class is to add an extra singlet with a new
(undetermined) Yukawa interaction with the Higgs fields. This is
sufficient to raise the Higgs mass from $130$ GeV up to about $150$ GeV
with fine-tuning comparable to the MSSM. But, any further upward push
on the Higgs mass causes the new Yukawa coupling to blow up at a scale
below the gauge coupling unification scale.

In fact, incorporating a significantly heavier Higgs mass into
beyond-the-MSSM models invariably leads to some type of strong
coupling behavior, such as a Landau pole in the Next-to-Minimal
Supersymmetric Standard Model (NMSSM), e.g.\ \cite{Haber:1986gz}, or
simply a low UV cutoff, e.g.\ \cite{Casas:2003jx}. This has usually
been viewed negatively because it ruins the UV completeness of
supersymmetric theories, and eliminates connections to attractive
theoretical ideas such as grand unification and string theory. The
usual lore is that strong coupling should be avoided, yielding an
upper bound on the mass of the Higgs.

We propose a radical revision of the usual lore and allow the Higgs
sector to become strongly coupled at an intermediate scale.  At the
scale of strong coupling the Higgs fields reveal their composite
nature and are in fact mesons of a confining theory in the UV\@. This
theory is renormalizable, asymptotically free, and thus UV complete.
The IR and UV dynamics are completely under control thanks to the
improved knowledge of strongly coupled supersymmetric gauge theories
\cite{Intriligator:1995au} that is reliable even when soft
supersymmetry breaking is added as a small perturbation on the strong
dynamics \cite{Arkani-Hamed:1998wc,Luty:1999qc}. We draw significant
inspiration from recent proposals to fuse supersymmetry with
technicolor \cite{Luty:2000fj,Murayama:2003ag}, and indeed in our
model electroweak symmetry is broken dynamically.  However, we have
physical (composite) Higgs fields in the low energy effective theory
with no \emph{a priori} restriction on the scale of strong coupling,
reminiscent of the older non-supersymmetric composite Higgs models
\cite{composite} (but without the associated fine-tuning problems).

The outline of this paper is as follows. We first discuss the
supersymmetric little hierarchy problem in Sec.~\ref{sec:little}, and
emphasize that a heavier Higgs mass easily solves this problem. We
then construct our supersymmetric composite Higgs theory in
Sec.~\ref{sec:composite}. Our basic framework is a three-flavor
$SU(2)_{H}$ theory that s-confines, resulting in a low energy
effective Lagrangian containing a dynamically generated superpotential
of composite mesons. By introducing a mass for one flavor below the
compositeness scale, we show that the mesons acquire expectation
values that break electroweak symmetry at a scale that is tunable
through this mass parameter. In Sec.~\ref{sec:scales} we demonstrate
how the various energy scales can be naturally obtained from the
supersymmetry breaking, and in Sec.~\ref{sec:fermion-masses} we show
how fermion masses and mixings can be incorporated. We calculate the
scalar spectrum in Sec.~\ref{sec:spectrum}. In Sec.~\ref{sec:pheno} we
briefly comment on the new phenomenology of this model, emphasizing
the unusual scalar spectrum. Sec.~\ref{sec:unification} explains how
gauge coupling unification can be preserved. Finally, we conclude with
a discussion in Sec.~\ref{sec:discussions}.

\section{Supersymmetric little hierarchy problem}
\label{sec:little}

In this section we define the supersymmetric little hierarchy problem
and propose a simple but unconventional way out. The problem is that
the conventional supersymmetric theories are increasingly fine-tuned
since the Higgs boson and/or charginos have not yet been discovered.
We point out that a composite Higgs will solve this problem easily
once a suitable UV completion is found.

The MSSM provides a simple way to understand electroweak symmetry
breaking (see, \emph{e.g.}, \cite{Martin:1997ns} for a review). In the
supersymmetric limit, electroweak symmetry is not broken. Therefore,
electroweak breaking is solely due to the soft supersymmetry breaking
effects. It arises from the renormalization of the up-type Higgs soft
mass-squared that is driven negative by the top Yukawa coupling. At
the one-loop approximation, one finds
\begin{equation}
  \label{toploop}
  \Delta m_{H_u}^2 \sim - 12 \frac{h_t^2}{16\pi^2} m_{\tilde{t}}^2 \log
  \frac{M_{UV}}{\mu_{IR}},
\end{equation}
where $M_{UV}$ ($\mu_{IR}$) is the UV (IR) cutoff and $h_t$ is the top
Yukawa coupling. Even with the universal boundary condition
$m_{\tilde{t}} = m_{H_u}$, it is easy to see that a large logarithm
between the weak scale and, say, the GUT-scale makes $m^2_{H_u}$
negative.  Assuming the supersymmetric mass $\mu$ for the Higgs
doublets is smaller in magnitude than $m_{H_u}$, electroweak symmetry
is broken.  This so-called radiative breaking of the electroweak
symmetry is a very nice feature of the MSSM\@.

However, the phenomenological situation is forcing some degree of
fine-tuning on the MSSM in the following fashion. First of all, the
Higgs quartic coupling is given only by $D$-terms that are determined
by the electroweak gauge couplings
\begin{equation}
  V_D = \frac{g^2 + g^{\prime 2}}{8} (|H_u|^2 - |H_d|^2)^2. 
\end{equation}
This implies the natural scale for the Higgs boson mass is $m_Z$, and
indeed there is a well-known tree-level upper limit on the lightest
Higgs mass that is precisely $m_Z$. The only way to increase the Higgs
mass is by using the $O(h_t^4)$ radiative correction to the Higgs
quartic coupling. The approximate formula valid for a moderate
$\tan\beta$ is
\begin{equation}
\label{MSSMhiggs}
  m_{h^0}^2 \simeq m_Z^2 + \frac{3}{4\pi^2} h_t^4 v^2 
  \log \frac{m_{\tilde{t}_1}m_{\tilde{t}_2}}{m_t^2}.
\end{equation}
Here, $v = 174$~GeV\@. Because the Higgs boson has not been found up
to 115~GeV, this implies $m_{\tilde{t}} \gtrsim 500$~GeV\@. On the
other hand, the minimization of the scalar potential leads to
\begin{equation}
  \label{Zmass}
  \frac{1}{2} m_Z^2 = - \mu^2 
  - \frac{m_{H_u}^2 \tan^2 \beta - m_{H_d}^2}{\tan^2 \beta -1}
  \simeq - \mu^2 - m_{H_u}^2,
\end{equation}
again for moderate $\tan\beta$. Therefore we need to fine-tune the
bare $m_{H_u}^2$ and/or $\mu$ against the radiative correction in
Eq.~(\ref{toploop}) at the level of
\begin{equation}
  \frac{|\Delta m_{H_u}^2|}{m_Z^2/2} \sim  4.8 
  \left( \frac{m_{\tilde{t}}}{\rm 500~GeV} \right)^2
  \log \frac{M_{UV}}{\mu_{IR}}.
\end{equation}
Even for a low UV scale of $M_{UV} = 100$~TeV, this already requires a
fine-tuning of 3\%.

In addition, the null results from searches for charginos at LEP-II
gives a lower bound $M_2 \gtrsim 100$~GeV\@.  Assuming a GUT relation
among the gaugino masses, this implies $M_3 \gtrsim
350$~GeV\@. Because $M_3$ feeds into $m_{\tilde{t}}$ through
renormalization group evolution, this then feeds into $m_{H_u}^2$,
aggravating the situation. Moreover, the MSSM potential is rather
delicate due to the possible instability along the $D$-flat direction
$H_u = H_d$.

The situation would clearly be better if the tree-level Higgs mass
could be raised above the LEP bound. Modifying Eq.~(\ref{Zmass}),
however, necessarily involves additional contributions to the Higgs
potential that are not related to the SM gauge couplings.
Furthermore, reducing the need for (s)top contributions to electroweak
symmetry breaking and the Higgs mass, Eqs.~(\ref{toploop}) and
(\ref{MSSMhiggs}) respectively, may help reduce the fine tuning
required. We will see that the Fat Higgs model we propose in this
paper achieves both of these aims.

The simplest extension of the MSSM that raises the tree-level Higgs
mass is the NMSSM.  In the NMSSM the $\mu$ term is replaced by the
superpotential
\begin{equation}
  W = \lambda N H_u H_d - \frac{k}{3} N^3
\end{equation}
where $N$ is neutral under the SM and $\lambda$ is undetermined.  The
Higgs quartic coupling therefore depends on $\lambda$ as well as the
gauge couplings, potentially allowing for a much higher Higgs
mass. Increasing the Higgs mass requires a large $\lambda$. This
coupling renormalizes \emph{upward} with increasing energy, eventually
encountering a Landau pole. At this scale the perturbative description
breaks down and the theory is no longer UV complete. To avoid this
problem, it is customary to impose the requirement that all coupling
constants, and in particular $\lambda$, remain perturbative up to the
gauge coupling unification scale. This places an upper bound on
$\lambda$ leading to an upper bound on the lightest Higgs mass of
about $150$~GeV
\cite{Haber:1986gz,Drees:tp,Espinosa:1991gr,Espinosa:1992hp,Cvetic:1997ky}.
Adding more matter fields can relax the bound somewhat, but not much
\cite{Moroi:1991mg,Kane:1992kq,Moroi:1992zk}.  Even for extensions of
the NMSSM with Higgs fields in other representations this bound is
relaxed to at most $m_h \lesssim 200$~GeV \cite{Espinosa:1998re}. This
is the basis for the lore that the lightest Higgs mass cannot be much
higher than in the MSSM\@.\footnote{In~\cite{Giudice:1998dj} a heavier
Higgs mass was claimed possible if a stop bound state condenses due to
a strong trilinear coupling. In~\cite{Batra:2003nj} a new moderately
strong gauge interaction was used to enhance the Higgs quartic
coupling, assuming a large supersymmetry breaking of about 7 TeV in a
part of the theory.}

This is the supersymmetric little hierarchy problem we intend to
solve. We are seeking a theory where the Higgs mass is allowed to be
much larger than $m_Z$ at tree-level. It is clear that the heart of
the problem in the MSSM is that the quartic coupling is determined by
the gauge couplings plus radiative corrections.  This is the ultimate
source of the tension between the stop masses and the lightest Higgs
mass. If the tree-level Higgs mass can be higher there is no need to
rely on the radiative corrections from the top-stop sector and
therefore stops can be light. In fact, if all superpartners are around
200--450~GeV, the natural scale for Higgs soft mass is also around the
same scale, and there is no fine-tuning.

In this paper we employ exact results in supersymmetric gauge theories
to UV complete a strongly coupled Higgs sector. In our model the Higgs
fields are composite bound states of fundamental fields charged under
a new, strongly coupled gauge theory.  The supersymmetric strong
dynamics drive electroweak symmetry breaking.  The effective theory of
the Higgs composites is a variant of the NMSSM with an arbitrarily
strong quartic coupling.  Furthermore, unlike the MSSM, the modified
potential has no flat directions that may cause an instability.

\section{The Fat Higgs}
\label{sec:composite}

First we describe the dynamics that leads to electroweak symmetry
breaking with composite ``fat'' Higgs fields.  The model is an $N=1$
supersymmetric $SU(2)$ gauge theory with six doublets, $T^1,\ldots,
T^6$. They carry the quantum numbers given in Table~\ref{qn-table}.
\begin{table}
\begin{center}
\begin{tabular}{l|ccccc}
Superfields & $SU(2)_L$ & $SU(2)_H$ & $SU(2)_R$ & $SU(2)_g$ & $U(1)_R$ \\
\hline\hline 
$(T^1, T^2) \equiv T$ & $\mathbf{2}$ & $\mathbf{2}$ & $\mathbf{1}$&
$\mathbf{1}$ & $0$ \\
$(T^3, T^4)$ & $\mathbf{1}$ & $\mathbf{2}$ & $\mathbf{2}$ &
$\mathbf{1}$ & $0$ \\ 
$(T^5, T^6)$ & $\mathbf{1}$ & $\mathbf{2}$ & $\mathbf{1}$ &
$\mathbf{2}$& $1$ \\ \hline
$P$ & $\mathbf{2}$ & $\mathbf{1}$ & $\mathbf{1}$ & $\mathbf{2}$ & $1$ \\ 
$Q$ & $\mathbf{1}$ & $\mathbf{1}$ & $\mathbf{2}$ & $\mathbf{2}$ &
$1$ \\ \hline 
$S$ & $\mathbf{1}$ & $\mathbf{1}$ & $\mathbf{1}$& $\mathbf{1}$ & $2$ \\
$S'$ & $\mathbf{1}$ & $\mathbf{1}$ & $\mathbf{1}$& $\mathbf{1}$ & $2$ 
\end{tabular}
\end{center}
\caption{Field content under $SU(2)_L$ $\times$ $SU(2)_H$ gauge and 
  $SU(2)_R \times SU(2)_g \times U(1)_R$ global symmetries. The
  $U(1)_Y$ subgroup of $SU(2)_R$ is gauged.}
\label{qn-table}
\end{table}

The tree-level superpotential consists of several terms,
\begin{equation}
W = W_1 + W_2 + W_3,
\end{equation}
where
\begin{eqnarray}
  W_1 &=& y S T^1 T^2 + y S' T^3 T^4 \\
  W_2 &=& - m T^5 T^6 \\
  W_3 &=& y (T^1, T^2) P \left(
    \begin{array}{c}
      T^5 \\ T^6
    \end{array} \right) +
  y (T^3, T^4) Q \left(
    \begin{array}{c}
      T^5 \\ T^6
    \end{array} \right).
\end{eqnarray}
The singlet fields $S$ and $S'$ in $W_1$ are necessary to ensure that
electroweak symmetry is indeed broken. $W_2$ is simply a mass term for
the fifth and sixth doublets. The mass parameter $m$ controls the
separation between the electroweak breaking scale and the
compositeness scale, as we will show. Finally, $W_3$ contains the
fields $P$ and $Q$ which are two-by-two matrices that transform as
doublets under $SU(2)_{L}$ or $SU(2)_{R}$ and also a global
$SU(2)_g$. They are present simply to marry off certain ``spectator''
composite fields with a mass of order the compositeness scale. $W_3$
is optional, since these spectator composite fields also acquire
electroweak symmetry breaking masses. But the addition of the $P$ and
$Q$ fields with the above superpotential has the benefit of minimizing
the field content of the low energy effective theory, which we call
the Minimal Supersymmetric Fat Higgs Model.

Note also that the overall superpotential is natural if we assign
non-anomalous $U(1)_R$ charges as shown in Table~\ref{qn-table}. The
global symmetries still allow for linear terms in $S$ and $S'$ in the
superpotential that could be forbidden by an additional $Z_3$
symmetry. This also prevents tadpole diagrams for the singlets that
could destabilize the weak scale~\cite{Bagger:1993ji}. Mass terms for
the first four doublets, if present, could be eliminated by shifting
the fields $S$ and $S'$. Our model is not sensitive to the precise
Yukawa couplings in the superpotential, but for simplicity we take a
common $y$ that is assumed to have the bare value $y_0 \sim
\mathcal{O}(1)$.

$SU(2)_{H}$ becomes strong at a scale $\Lambda_{H}$.  The theory has
six doublets, so below $\Lambda_{H}$ it is described by meson
composites $M_{ij}=T^i T^j$, ($i,j=1, \ldots ,6$) with a dynamically
generated superpotential ${\rm Pf} M/\Lambda^3$. Together with the
tree-level terms,
\begin{eqnarray}
  W_{\it eff} &=& \frac{{\rm Pf} M}{\Lambda^3} - m M_{56}
  + y S M_{12} + y S' M_{34} \nonumber \\
& &{} + y P^{k,\alpha} M_{k,\alpha+4} + y Q^{l, \alpha} M_{l+2, \alpha+4}
\end{eqnarray}
where $k=1,2$ is an $SU(2)_L$ index contracted with $P$; $l=1,2$ is an
$SU(2)_R$ index contracted with $Q$; and $\alpha = 1,2$ is an $SU(2)_g$
index. In terms of the canonically normalized fields this becomes
\begin{eqnarray}
  W_{dyn} &=& \lambda ({\rm Pf}M - v_0^2 M_{56})
  + m_{\rm spect} \big( S M_{12} + S' M_{34} \nonumber \\
  & &  + P^{k,\alpha} M_{k,\alpha+4} + Q^{l, \alpha} M_{l+2, \alpha+4} \big).
\end{eqnarray}
where Naive Dimensional Analysis (NDA)~\cite{NDA}
suggests\footnote{Here, the parameters are defined at the scale
  $\Lambda_H$, and hence are not the bare ones. As we will see in the
  next section, $m \sim 4\pi m_0$, $y \sim 4\pi y_0$, and $\Lambda_H
  \sim m' \sim 4\pi m'_0$, due to the superconformal dynamics.}
\begin{eqnarray}
  v_0^2 &\sim& \frac{m\Lambda_{H}}{(4\pi)^2},\\
  m_{\rm spect} &\sim& y \frac{\Lambda_{H}}{4\pi},\\
  \lambda (\Lambda_{H}) &\sim& 4\pi.
\end{eqnarray}
The crucial observation is that the scale of electroweak symmetry
breaking, $v_0$, is generated dynamically and is controlled by the
value of the supersymmetric mass $m$.

It is useful to change the notation for the meson matrix to make the
role of different components clear:
\begin{equation}
  N = M_{56}, \,
  \left(
    \begin{array}{c}
      H_u^+ \\ H_u^0
    \end{array} \right)
  = \left(
    \begin{array}{c}
      M_{13} \\ M_{23}
    \end{array} \right), \,
  \left(
    \begin{array}{c}
      H_d^0 \\ H_d^-
    \end{array} \right)
  = \left(
    \begin{array}{c}
      M_{14} \\ M_{24}
    \end{array} \right),
\end{equation}
and all the other components of the meson matrix decouple near $\Lambda_{H}$.
These Higgs fields have the dynamically generated superpotential
\begin{eqnarray}
  W &=& \lambda N (H_d H_u - v_0^2) \; .
\end{eqnarray}
The $H_d$ and $H_u$ doublets play the role of the MSSM Higgs doublets.
As advertised, this superpotential forces electroweak symmetry
breaking without relying on supersymmetry breaking effects.

The strong coupling $\lambda$ rapidly renormalizes to smaller values
as the energy scale is reduced. We can estimate this running by
neglecting corrections from gauge couplings and the top coupling. The
solution to the one-loop renormalization group equation
is\footnote{This formula assumes that the spectators decouple at scale
$\Lambda_H$. This is justified in the next section where the
superconformal dynamics enhances $y$ to $y \sim 4\pi y_0 \sim 4\pi$.
}
\begin{equation}
  \lambda^2(t) = \frac{2 \pi^2}{2 \pi^2\lambda^{-2}(0) + t}\, ,
\end{equation}
where $t \equiv \log (\Lambda_{H}/\mu)$ \emph{increases} towards the
infrared. Using the NDA estimate $\lambda(0) \sim 4\pi$, we find, for
example, $\lambda(4.5) \simeq 2$, practically independent of the initial
value. If $m$ is well below $\Lambda_{H}$, condensation occurs at the scale
$4\pi v_0 \sim (m \Lambda_{H})^{1/2} \ll \Lambda_{H}$ where the theory is
weakly coupled and therefore calculable.

This is one of our important results. In the infrared the theory
contains Higgs states with a weakly coupled, renormalizable
superpotential described by just two parameters, $\lambda$ and $v_0$.
This is a rather nontrivial result that depends on the specific choice
of an $SU(2)_H$ gauge theory with three flavors.  Other
choices are not so interesting.  For example, precision electroweak
constraints tend to severely restrict new gauge interactions beyond
the SM at the TeV scale, and thus theories with a dual magnetic
description are not good candidates. For an $SU(N_c)$ theory there is a
dynamically generated superpotential when $N_f = N_c + 1$, and
its renormalizability requires $N_f\leq 3$. This means that an $SU(2)$
gauge theory with three flavors is the unique choice for this purpose.

\section{Scales}
\label{sec:scales}

Phenomenologically, the scale of supersymmetry breaking soft masses
must be near the electroweak scale, $\lambda v_0 \sim m_{\rm SUSY}$,
because much larger SUSY-breaking would lead to fine-tuning, whereas
a much smaller SUSY-breaking scale would have
already been observed. Using the parameters of the UV theory, this
implies $m \Lambda_{H} \sim (4\pi m_{\rm SUSY})^2$. This coincidence
of scales is reminiscent of the $\mu$-problem in the MSSM\@. Here we
show that this can be naturally obtained by a combination of the
seesaw mechanism~\cite{seesaw} and the Giudice--Masiero mechanism
\cite{Giudice:1988yz}. This requires conventional
gravity-mediated supersymmetry breaking, which we assume for the
discussion in this section.

\begin{figure}[tc]
\centering \includegraphics[width=\columnwidth]{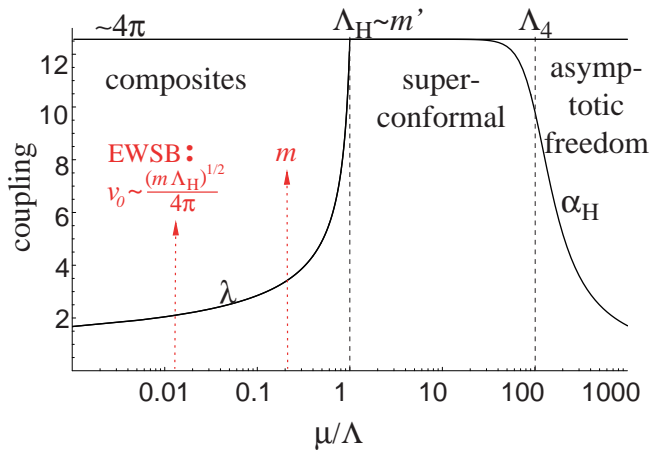}
\caption{The renormalization of the couplings in our Fat Higgs
model. The model becomes strong and nearly conformal at the scale
$\Lambda_4$, where $\alpha_{H}$ nears $4\pi$. The conformal
invariance is broken by the mass of the extra doublet, $m'$, which
makes the theory confine at $\Lambda_{H} \sim m'$. Below this scale
the effective theory description becomes one of meson composites with
a coupling $\lambda$ that quickly renormalizes down to
${\cal O}(1)$. When $4\pi v_0\ll \Lambda_{H}$ the mesons condense at
weak coupling and the theory is calculable.}
\label{fig:RGE2}
\end{figure}

The simplest way to relate $\Lambda_{H}$ to other scales is
to employ a superconformal theory where the gauge coupling
remains constant over many decades in energy. We introduce two extra
doublets $T^7$ and $T^8$ to $SU(2)_{H}$ for this purpose.
We assume $T^7$ and $T^8$ transform as a vector-like pair under other
symmetries so that a supersymmetric mass term can be added to the
superpotential
\begin{equation}
  W = m' T^7 T^8 \; .
\end{equation}
This theory with $N_c=2$ and $N_f=4$ is in the superconformal window
\cite{Intriligator:1995au}. At some scale $\Lambda_4$ the $SU(2)_{H}$ gauge
coupling becomes strong and remains strong all the way down to $m'$, the
supersymmetric vector-like mass of the extra doublets. At the scale $m'$
the conformal symmetry is broken and $T^{7,8}$ may be integrated out. Below
this scale the theory confines and is effectively the three flavor model
discussed in the previous section. We therefore identify the strong coupling
scale $\Lambda_{H}$ with $m'$. The renormalization group evolution of the
couplings is schematically shown in Fig.~\ref{fig:RGE2}.

In addition to determining the scale $\Lambda_H$, the conformal
dynamics generate large anomalous dimensions which have the effect of
enhancing the couplings of the $T$ fields, and therefore also the
couplings of the composite Higgs fields.  The structure of the
superconformal algebra determines the anomalous dimensions exactly in
terms of the anomaly-free $R$-charges.  Running from the strong scale
$\Lambda_4$ down to the scale of conformal breaking $\Lambda_H$, the
wave function of the $T$'s is suppressed as
\begin{equation}
Z \sim \left( \frac{\Lambda_H}{\Lambda_4}\right)^{\gamma_*}
\end{equation}
where $\gamma_*=1/2$ is the anomalous dimension. Once the fields
are canonically normalized this leads to an \emph{enhancement} of their
couplings. For example, the effective mass $m'$ gets enhanced by a factor of  
\begin{eqnarray}
  \label{enhancement}
  \left( \frac{\Lambda_4}{\Lambda_H} \right)^{1/2}.
\end{eqnarray}
In the low energy theory, any operator that involves one Higgs field, such as
the top Yukawa, will be enhanced by a similar factor. 
Because the superconformal dynamics is likely to be upset by other
strong couplings, the largest enhancement factor we consider is $4\pi$.

The next task is to determine how $m$ of the right size can be generated. 
First, it is assumed that the heavier vector-like mass $m'$ is
unrelated to supersymmetry breaking and therefore arbitrary. The
scale for $m'$ is presumably set by other flavor symmetries, akin to
the right-handed neutrino mass which is protected by lepton number. However,
the symmetries may conspire to forbid a vector-like mass $m$ for 
the third flavor, analogous to the left-handed neutrino mass in the
neutrino mass matrix. 
For example, consider a simple $U(1)$ flavor symmetry of 
charge $+1$ ($-1$) for the third (fourth) flavor. The symmetry is broken 
by an order parameter of charge $+2$. Then $m'$ is allowed in the 
superpotential while $m$ is not. Nevertheless, mixing between the third 
and fourth flavors is allowed by the symmetries and originates from the
supersymmetry breaking due to the Giudice--Masiero mechanism.
Therefore, the form of the mass matrix for these flavors becomes
\begin{equation}
  \left( 
    \begin{array}{cc}
      0 & m_{\rm SUSY} \\
      m_{\rm SUSY} & m'
    \end{array} \right).
\end{equation}
The light eigenvalue is given by $m = m_{\rm SUSY}^2/m'$. After the
conformal dynamics enhances both $m$ and $m'$, we naturally obtain $m
m' \sim (4\pi m_{\rm SUSY})^2$ as desired.

\section{Fermion Masses}
\label{sec:fermion-masses}

In order to incorporate fermion masses, we
follow~\cite{Murayama:2003ag} by adding four additional chiral
multiplets that are singlets under $SU(2)_{H}$ but have the same
quantum numbers as the Higgs doublets $H_u$ and $H_d$ in the MSSM,
\begin{equation}
  \varphi_u, \bar{\varphi}_d ({\bf 1}, {\bf 2}, +\frac{1}{2}), \quad
  \varphi_d, \bar{\varphi}_u ({\bf 1}, {\bf 2}, -\frac{1}{2}).
\end{equation}
They have the superpotential
\begin{eqnarray}
  W_{f} &=& M_{f} (\varphi_u \bar{\varphi}_u + \bar{\varphi}_d
  \varphi_d) + \bar{\varphi}_d (T T^4)
  + \bar{\varphi}_u (T T^3) \nonumber \\
  &&+ h_u^{ij} Q_i u_j \varphi_u + h_d^{ij} Q_i d_j \varphi_d
  + h_e^{ij} L_i e_j \varphi_d.
\end{eqnarray}
where $M_f$ is the mass of $\varphi$ and $\bar{\varphi}$. 
The only flavor-violating couplings are the Yukawa couplings
$h_u^{ij}$, $h_d^{ij}$, $h_e^{ij}$. We assume $M_f\sim m'\sim
\Lambda_H$, possibly due to the same flavor symmetries that control
the size of $m'$.

Between $\Lambda_4$ and $\Lambda_H \simeq m'$ the superconformal
dynamics enhances the Yukawa couplings by $(\Lambda_4/\Lambda_H)^{1/2}
\sim 4\pi$, as described in the previous section.  After the
$\varphi$'s are integrated out, the effective dimension-5
superpotential is
\begin{eqnarray}
  W_{f} &=& \frac{4\pi}{M_{f}} \Big[ h_u^{ij} Q_i u_j (T T^3)
    + h_d^{ij} Q_i d_j (T T^4) \nonumber \\
& &{} + h_e^{ij} L_i e_j (T T^4) \Big].
\end{eqnarray}
Below the compositeness scale $\Lambda_H$, NDA specifies the
replacement $(T T^3) \rightarrow \Lambda_H H_u/4\pi$, $(T T^4)
\rightarrow \Lambda_H H_d/4\pi$.  Using $M_f\sim\Lambda_H$, 
the superpotential becomes
\begin{equation}
  W_{f} =   h_u^{ij} Q_i u_j H_u + h_d^{ij} Q_i d_j H_d
  + h_e^{ij} L_i e_j H_d.
\end{equation}

One may wonder if the Yukawa couplings are suppressed in the
low-energy theory due to the wavefunction renormalization of the Higgs
fields due to the strong coupling $\lambda$. Again using the one-loop
renormalization group equation for simplicity, we find
\begin{equation}
  h(t) = h(0) \left( \frac{\lambda(t)}{\lambda(0)} \right)^{1/4}. 
\end{equation}
For $\lambda(0) \sim  4\pi$ and $\lambda(t) \sim 2$,  we find that the
suppression  is only  60\%. Therefore  the mechanism  presented above
does yield sufficiently large Yukawa couplings.

\section{Higgs Mass Spectrum}
\label{sec:spectrum}

In this section the mass spectrum of the model is calculated. The
supersymmetric part of the Higgs potential is
\begin{equation}
  V_{\rm SUSY} = \lambda^2 |H_d H_u - v_0^2|^2 
  + \lambda^2 |N|^2 (|H_u|^2 + |H_d|^2),
  \label{pot1-eq}
\end{equation}
together with the $D$-term contributions that are familiar from the MSSM,
\begin{equation}
  V_D = \frac{g^2}{8} (H_u^\dagger \vec{\tau} H_u + H_d^\dagger
  \vec{\tau} H_d)^2 + \frac{g^{\prime 2}}{8} (|H_u|^2 - |H_d|^2)^2.
\end{equation}
Unlike the MSSM, electroweak symmetry breaking is caused by the
confining dynamics even in the absence of supersymmetry breaking.
Nevertheless, the potential also contains soft supersymmetry breaking
terms
\begin{eqnarray}
  V_{\it soft} &=& m_1^2 |H_d|^2 + m_2^2 |H_u|^2 + m_0^2 |N|^2
  \nonumber \\
  & & + (A \lambda N H_d H_u - C \lambda v_0^2 N + h.c.)
  \label{pot2-eq}
\end{eqnarray}
where $m_1, m_2, m_0, A, C \sim m_{\rm SUSY}$.

It is instructive to first look at a simple case where $m_1 = m_2 =
m_0$, $A=C=0$. In this case, we can define the ``Standard
Model-like Higgs'' $H = (H_u^0 + H_d^0)/\sqrt{2}$ whose potential is
simply
\begin{eqnarray}
  V &=& \frac{1}{4} \lambda^2 |H^2 - 2v_0^2|^2 + m_0^2 |H|^2
  \nonumber \\
  &=& \frac{1}{4}\lambda^2 |H^2|^2 - (\lambda^2 v_0^2 - m_0^2) |H|^2
  + {\rm const}.
  \label{pot3-eq}
\end{eqnarray}
This is no different from the potential in the minimal Standard Model.
It is clear that electroweak symmetry is broken so long as $\lambda^2
v_0^2 > m_0^2$. The vacuum expectation value is $v = \langle
H \rangle = \sqrt{2(v_0^2 - m_0^2/\lambda^2)}$, and the mass
of the Higgs boson is $\lambda v$. There is an exact custodial $SU(2)$
symmetry in the Higgs sector.

For a more general set of parameters it becomes difficult to solve for
the vacuum analytically.  
We take advantage of the (small) hierarchy
\begin{equation}
  \lambda v_0 \sim m_{1,2} \gg gv_0, g'v_0,
\end{equation}
which allows us to drop the MSSM $D$-terms.  
If $m_1/m_2$ is
very large, the quartic term in the potential is dominated by the
$D$-term and cannot be ignored.  Our solution for the vacuum state 
applies for a moderate ratio $m_1/m_2$ where both Higgs masses $m_{1,2}$ 
are much larger than $m_Z$.

The fact that the Higgs quartic coupling of the MSSM is
negligible compared to that coming from the strong dynamics illustrates
that our model manifestly solves the supersymmetric little hierarchy problem, 
as anticipated in Section~\ref{sec:little}. For simplicity we also 
set $A=C=0$ for most of the discussions below. With these approximations, 
the ground state is
\begin{eqnarray}
  H_u^0 &=& v_0\sqrt{1-\frac{m_1 m_2}{\lambda^2v_0^2}}\sqrt{\frac{m_1}{m_2}} ,
  \\
  H_d^0 &=& v_0\sqrt{1-\frac{m_1 m_2}{\lambda^2v_0^2}}\sqrt{\frac{m_2}{m_1}} ,
  \\
  N &=& 0
\end{eqnarray}
up to corrections of order $m_Z^2/m_{1,2}^2$. 
To leading order in $A$ and $C$ we find that $N$ no longer vanishes,
\begin{equation} \label{eqn:N}
  N = \frac{m_1 m_2 ((-A+C) \lambda^2 v_0^2 + A m_1 m_2)}{\lambda
  ((\lambda^2 v_0^2 - m_1 m_2) (m_1^2+m_2^2) + m_1 m_2 m_0^2)},
\end{equation}
while shifts to $H_u$ and $H_d$ are only $O(A^2,AC,C^2)$. This
demonstrates that a $\mu$-term is naturally
generated, giving a mass of order $m_{\rm SUSY}$ to the Higgsinos.

Our vacuum solution was obtained by assuming that $m_{0,1,2}^2 > 0$, 
and thus arises from dynamical (as opposed to radiative) breaking of 
electroweak symmetry.  Nevertheless, we expect a stable vacuum with 
electroweak symmetry breaking even if some or all of (mass)$^2$ are 
negative because our potential Eq.~(\ref{pot1-eq})-(\ref{pot3-eq}) 
is bounded from below, unlike in the MSSM where there is a possible 
instability along the $D$-flat direction.  We leave such cases for a 
future study.

It is rather convenient to define
\begin{eqnarray}
m_s &=& \sqrt{m_1 m_2} \\
\tan\beta &\equiv& \frac{\langle H_u^0 \rangle}{\langle H_d^0 \rangle} 
  = \frac{m_1}{m_2} \; ,
\end{eqnarray}
and then the electroweak breaking scale $v \simeq 174$~GeV is 
fixed in terms of the parameters of the model,
\begin{equation}
  \label{Wmass}
  v^2 = 2 \frac{\lambda^2 v_0^2 - m_s^2}{\lambda^2 \sin 2\beta} \; ,
\end{equation}
with the usual $W$ and $Z$ masses
\begin{equation}
m_W^2 = \frac{1}{2}g^2v^2 \qquad \mbox{and} \qquad 
m_Z^2 = \frac{1}{2}(g^2+g'^2)v^2.
\end{equation}

The charged Higgs states have mass
\begin{equation}
  m^2_{H^\pm} = \frac{2 m_s^2}{\sin 2\beta}.
\end{equation}
The singlet state $N$ has both scalar and pseudo-scalar states that are
degenerate (within our simplifying assumption of $A=C=0$),
\begin{equation}
  m^2_{N_1} = m^2_{N_2} = \lambda^2 v^2 + m_0^2\, ,
\end{equation}
while the pseudo-scalar from the Higgs doublets has mass
\begin{equation}
  m^2_{A^0} = \lambda^2 v^2 + m_{H^\pm}^2.
\end{equation}
The neutral scalars from the doublets have a mass matrix
\begin{equation}
  \left(  \begin{array}{cc}
      (\lambda^2 v^2 + m_{H^\pm}^2) \cos^2 \beta &
      (\lambda^2 v^2 - m_{H^\pm}^2) \sin \beta \cos \beta\\
      (\lambda^2 v^2 - m_{H^\pm}^2) \sin \beta \cos \beta &
      (\lambda^2 v^2 + m_{H^\pm}^2) \sin^2 \beta
    \end{array}
  \right),
\end{equation}
where the upper (lower) components correspond to the $h_u^0$ ($h_d^0$)
defined by the expansion $H_{u,d}^0 = \langle H_{u,d}^0 \rangle +
h^0_{u,d}/\sqrt{2}$. This mass matrix leads to the eigenvalues
\begin{eqnarray}
\label{higgs-spectrum}
  m^2_{H^0}&=& \frac{\lambda^2 v^2 + m_{H^\pm}^2 + X}{2}\, ,
  \\
  m^2_{h^0}&=&\frac{\lambda^2 v^2 + m_{H^\pm}^2 - X}{2}\, ,
\end{eqnarray}
where 
\begin{equation} \label{eqn:x}
  X = \sqrt{  (\lambda^2 v^2 + m_{H^\pm}^2)^2
            - 4 \lambda^2 v^2 m_{H^\pm}^2 \sin^2 2\beta}\ .
\end{equation}
They are given in terms of the
gauge eigenstates by
\begin{equation}
  \left( \begin{array}{c} h^0 \\ H^0 \end{array} \right)  = 
  \left(\begin{array}{cc} \cos\alpha & -\sin\alpha \\ \sin\alpha
      & \cos\alpha \end{array} \right)
  \left( \begin{array}{c} h_u^0
      \\ h_d^0 \end{array}
  \right)\, ,
\end{equation}
where $\alpha$ is the mixing angle that in our model is 
\begin{equation}
  \tan\alpha = \frac{(\lambda^2 v^2+ m_{H^\pm}^2)\cos2\beta + 
    X}{(\lambda^2 v^2 - m_{H^\pm}^2)\sin2\beta}.
\end{equation}
The above expressions will receive corrections from the
$D$-terms at the level of $m_Z^2/m_{\rm SUSY}^2$ and $m_Z^2/\lambda^2
v^2$.

Here we highlight some of the interesting general features of this
spectrum. The model contains singlet scalar and pseudo-scalar states
$N_1$ and $N_2$ that are not present in the MSSM. In addition, the
pseudo-scalar Higgs $A^0$ is always heavier than the charged Higgs,
which is the opposite of the MSSM. When studying the other states, it
is useful to consider two limiting cases, $\lambda v \ll m_{H^\pm}$
and $\lambda v \gg m_{H^\pm}$. In the first case, $\tan\alpha
\rightarrow -\cot\beta$, and hence the lighter eigenstate ``aligns''
with the vacuum. In other words, the lighter neutral Higgs $h^0$ is
Standard Model-like, while the heavier state $H^0$ does not have
couplings to $ZZ$ or $W^+ W^-$. It forms a custodial $SU(2)$ triplet
with the charged Higgs, $(H^+, H^0, H^-)$. In the second case,
$\tan\alpha \rightarrow \cot\beta$. If $\tan\beta=1$ they are still
aligned, and the {\it heavier}\/ neutral Higgs $H^0$ is Standard
Model-like, and the custodial $SU(2)$ triplet is $(H^+, h^0,
H^-)$. For a general $\tan\beta$, however, the eigenstates and the
vacuum are not aligned, and both neutral states contain the Standard
Model-like Higgs state. In particular, the mixture is maximal if
$\beta = 3\pi/8$ ($\tan\beta = 2.414$).

\section{Phenomenology}
\label{sec:pheno}

Since the theory is supersymmetric, we expect superpartners just like
in the ordinary MSSM\@. However, there are important distinctions
between our Fat Higgs model and the MSSM, especially in the Higgs
spectrum. This is the main issue we discuss in this section.

\subsection{Spectrum}

To study the phenomenology of the Minimal Supersymmetric Fat Higgs
model, we pick three points in the parameter space:
\begin{equation}
  \begin{tabular}{@{\extracolsep{.3cm}}c|ccc}
    & $\lambda$ & $\tan\beta$ & $m_s$ ({\rm GeV}) \\ \hline
    I & 3 & 2 & 400  \\
    II & 2 & 2 & 200  \\
    III & 2 & 1 & 200
  \end{tabular}
\end{equation}
$m_0$ is chosen to be the same as $m_s$ for simplicity; changing $m_0$
merely brings the mass of the $N_{1,2}$ states up and down independent 
of the rest of
the spectrum (within our simplifying assumption $A=C=0$).  

\begin{figure}[tc]
  \centering \includegraphics[width=\columnwidth]{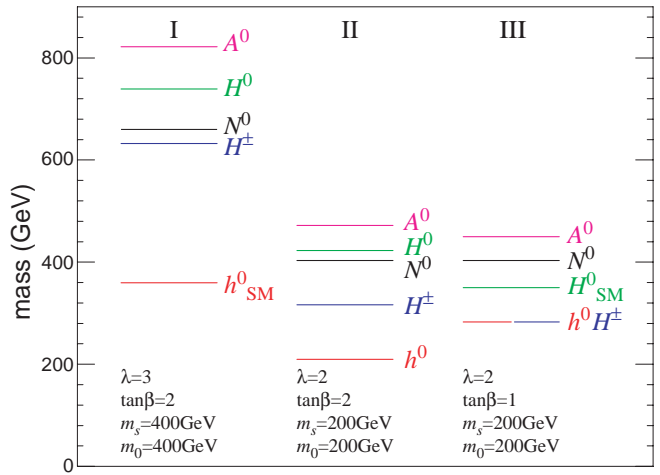}
  \caption{Sample Higgs spectra in our Fat Higgs model. In Spectrum I
  the SM-like Higgs is dominantly $h^0$ (89\%), whereas in Spectrum
  III the SM-like Higgs is purely $H^0$.}  \label{higgsspectrum-fig}
\end{figure}

Spectrum I corresponds to the case $m_{H^\pm} > \lambda v$ where
the lightest neutral Higgs is Standard Model-like, while the heavier
neutral Higgs $H^0$ forms a triplet with the charged Higgses $H^\pm$
under the approximate custodial $SU(2)$ symmetry.  The pseudo-scalar
Higgs $A^0$ is heavier than both of them.  It resembles the spectrum
in the MSSM when $A^0$ is heavy, but the relative ordering of the
heavy Higgs states is quite different.

Spectrum II has a smaller supersymmetry breaking scale, and both $h^0$
and $H^0$ have significant Standard Model-like Higgs content,
approximately 75\% and 25\%, respectively.  Such a large mixing is
unusual in the MSSM when the masses are this different.

Spectrum III is the most unconventional of all.  Because of the
exact custodial $SU(2)$, the triplet $h^0$ and $H^\pm$ are degenerate,
and they do not contain the Standard Model-like Higgs component.  On
the other hand, the heavier neutral Higgs $H^0$ is Standard
Model-like.  The pseudo-scalar Higgs is even heavier.

\subsection{Electroweak constraints}

It is well known that as the SM-like Higgs mass is raised above about
$250$ GeV the SM without new physics is increasingly disfavored by
electroweak precision data. In our Fat Higgs model, however, there are
several contributions to electroweak observables that are around the
same size as the one from a heavier SM-like Higgs. We have calculated
the contribution of the Higgs states to the electroweak parameters
$S$, $T$ \cite{PT}. The analytical results are the same as the MSSM,
and we present formulae in Appendix~\ref{sec:higgs} for
completeness.

\begin{figure}[tc]
  \centering \includegraphics[width=0.5\textwidth]{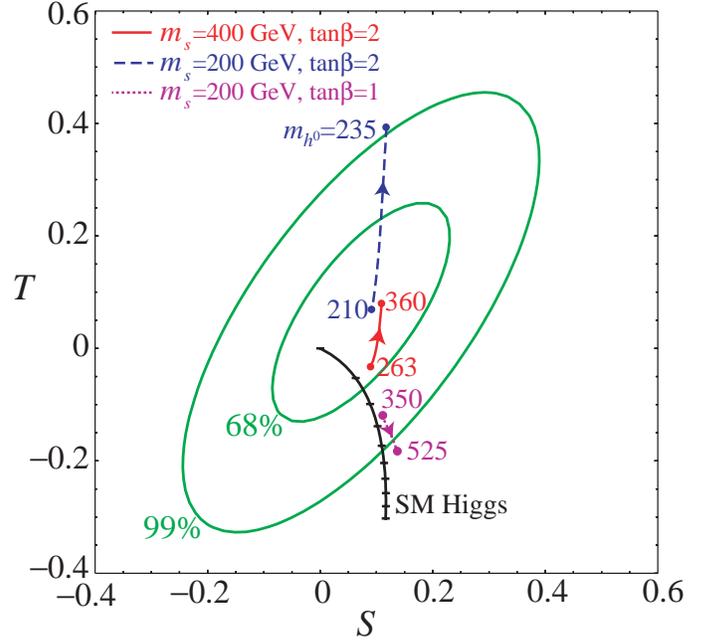}
  \caption{Constraints on $S$ and $T$ parameters from precision
  electroweak data at 68\% and 99\% confidence levels. The plot
  assumes $U=0$. Contributions of the Fat Higgs model to $S$ and $T$
  are shown along three trajectories where $\lambda$ is varied from 2
  to 3 in the direction of the arrow. The endpoints are labeled with
  the mass (in GeV) of the lighter component of the SM-like Higgs. For
  comparison, the black line shows the contributions to $S$ and $T$
  for the Standard Model with various Higgs masses between $100$ GeV
  and 1 TeV in increments of 100 GeV.}  \label{fig:ST}
\end{figure}

We find that the model is consistent with the experimentally allowed
region in the $S$-$T$ plane, described in Appendix~\ref{sec:ST},
with no fine-tuning of
model parameters. As an example of this, we present the $S$ and $T$
contributions for three trajectories in parameter space in
Fig.~\ref{fig:ST}.

In two trajectories, the coupling $\lambda(v_0)$ is varied between 2 and
3 for $\tan\beta=2$ and for two
different SUSY breaking scales. The mass of the lightest component of 
the SM Higgs is shown at the endpoints of each trajectory. Spectrum I of
Fig.~\ref{higgsspectrum-fig} is at the top of the solid trajectory and
Spectrum II is at the bottom of the dashed one. Note that these two
spectra are in excellent agreement with electroweak constraints despite
the heavy Higgs masses of 360 and 210 GeV. Furthermore, we find that the
constraints for $S$ and $T$ are easily satisfied for a significant
range of the model parameters, as the two trajectories demonstrate.

\begin{figure}[tc]
  \centering
  \includegraphics[width=0.5\textwidth]{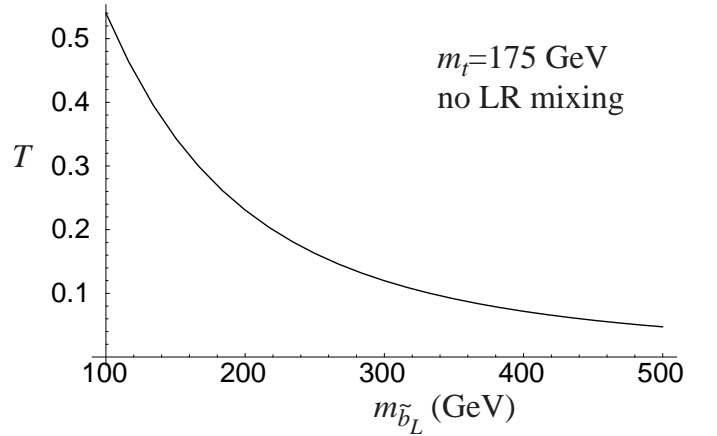}
  \caption{Contribution of the stop-sbottom sector to the
  $T$-parameter.}
  \label{fig:stop} 
\end{figure}

In the third (dotted) trajectory, $\lambda(v_0)$ is varied between 2
and 3 for $\tan\beta=1$. Spectrum III in Fig.~\ref{higgsspectrum-fig}
is at the top of this line. This trajectory lies mostly within the
99\% CL contour despite the unconventional Higgs spectrum.  However,
it is well known that a stop-sbottom splitting ($m_{\tilde{t}_L}^2 =
m_{\tilde{b}_L}^2 + m_t^2$) may contribute significantly to $T$ (see
Fig.~\ref{fig:stop}). This contribution may easily be $0.1$--$0.5$
(e.g.~\cite{Drees:1990dx}) that would bring these points back into the
68\% CL ellipse.  The size of this contribution depends on the masses
in the stop-sbottom sytem which can generically be different from the
SUSY breaking masses in the Higgs sector.

Nevertheless we stress that even with a negligible contribution to $T$
from the stop-sbottom sector, the Higgs mass can be hundreds of GeV,
as shown by trajectories I and II, and yet still stay well within the
precision electroweak contours.

\subsection{$b \rightarrow s\gamma$ constraint}

Another constraint on our model comes from $b \rightarrow s\gamma$
transitions mediated by charged Higgs bosons. Considering only
the charged Higgs/top quark contribution, the constraint on the
charged Higgs mass is $m_{H^\pm} > 350$ GeV at 99\% CL
\cite{Gambino:2001ew}. There are two ways this constraint could be
satisfied. The first is to simply raise the charged Higgs mass above
the bound. The second is the well-known possibility that the
charged Higgs contribution cancels against the chargino-stop
contribution, and therefore allows a lighter charged
Higgs~\cite{bsgcancellation}. This of course depends on the specific
model and parameters for supersymmetry breaking.

\subsection{Search strategies}

The neutral Higgs scalars $h^0$ and $H^0$ each may have a significant component
of the Standard Model-like Higgs boson and can thus be discovered by the
standard search methods, in particular the ``gold-plated'' signal
$h_{\rm SM} \rightarrow ZZ \rightarrow 4\ell$ at the LHC\@. However,
the decay modes $H^0 \rightarrow h^0 h^0, H^+ H^-$ may also be
open, and their partial decay widths are all
comparable and proportional to $\lambda^2 m_{H^0}$.

In the strict custodial $SU(2)$ limit, when $\tan\beta =1$, the triplet
Higgses are produced only in pairs. In particular, there is no
production process of the neutral state by itself, such as $e^+ e^-
\rightarrow Z h$ or $u\bar{d} \rightarrow W^+ h$, when $h$ does not
contain the Standard Model-like Higgs component. Nevertheless, they
do have the top Yukawa coupling and thus can be produced from the
gluon fusion at the LHC\@. Their decays depend very sensitively on
the superpartner spectrum and the Higgs spectrum.

In order to positively establish that our model correctly describes
the Higgs sector, numerous other measurements will be needed: the
complete mass spectrum, branching fractions, and Higgs self-couplings.
For this purpose, an $e^+ e^-$ Linear Collider would be a great asset.

\begin{figure}[tc]
\centering \includegraphics[width=\columnwidth]{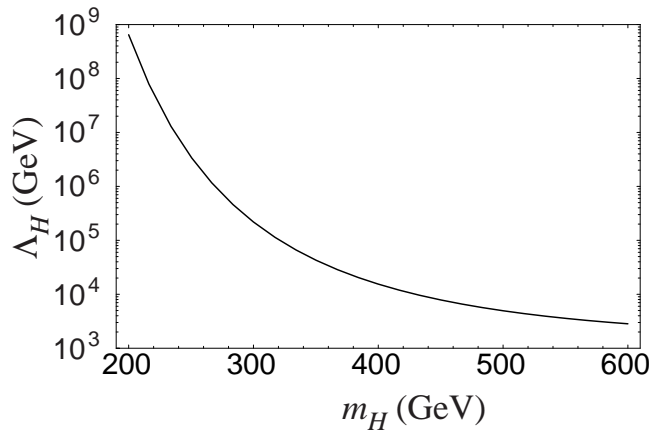}
\caption{The compositeness scale is shown as a function of the 
Higgs mass, fixing $\tan\beta = 1$. This was determined by finding the
scale $\Lambda_H$ where $\lambda=4\pi$, using one-loop renormalization
group evolution.}
\label{fig:Landau}
\end{figure}

Once the Higgs mass is measured, we know its quartic coupling
$\lambda$. Because of the renormalization group evolution, a lower
compositeness scale corresponds to a heavier Higgs mass.
This is shown in Fig.~\ref{fig:Landau}.
The limit of the low compositeness scale is of course of special
interest, where we may have direct access to the composite dynamics of
the Higgs. Because this limit has its own special issues, we will
defer the discussion of this case, the ``Fattest Higgs'', to a
separate paper.\footnote{The limit $m\rightarrow m'$ is identical to
Technicolorful Supersymmetry~\cite{Murayama:2003ag} except for the
presence of the $P$ and $Q$ fields. Our Fat Higgs model is hence an
``analytic continuation'' of Technicolorful Supersymmetry.}

\subsection{Cosmology}

The NMSSM is known to have a cosmological problem due to the
spontaneous breaking of its $Z_3$ symmetry that produces domain walls
(see~\cite{Nevzorov:2001ny} for a recent discussion on this issue).
Interestingly, our Fat Higgs model does not contain such a symmetry
and is free from the domain wall problem.  The $Z_3$ symmetry used to
forbid the linear terms in $S$ and $S'$ acts on $T^{1,2}$ with charge
$+1$ and $T^{3,4}$ with charge $-1$, and hence all Higgs fields $N$,
$H_{u,d}$, quarks, and leptons are neutral under this $Z_3$.  On the
other hand, $R$-parity can be imposed consistently and thus
the lightest supersymmetric particle is a candidate for cold dark
matter.  

Finally, note that the charge assignments given in
Table~\ref{qn-table} for $T^{5,6}$ and the $P$'s and $Q$'s lead to
fractionally charged spectators with electric charge $\pm 1/2$. The
lightest stable one does not decay, and this could lead to problems in
early universe cosmology. There are several options: The first is to
simply assume that these particles are not produced after reheating by
restricting the reheat temperature to be much lower than their mass. A
second possibility is to change the charge assignments of $T^{5,6}$ so
that they carry $\pm 1/2$ hypercharge, and therefore all the low
energy composites have integral charge. We will see below, however,
that leaving the charge assignment as given in Table~\ref{qn-table}
allows the simplest interpretation of gauge coupling unification in
the model.

\section{Unification}
\label{sec:unification}

In this section we complete the discussion of our Fat Higgs model
by showing that gauge coupling unification can be easily preserved
despite the composite nature of the Higgs fields and the strong coupling. 
The effective theory below the compositeness scale (and below 
$M_f$ and $m_{\rm spect}$) has the same matter content as the NMSSM,
thus the gauge couplings run exactly like the MSSM gauge 
couplings until the compositeness scale is reached.
That the couplings can unify above the compositeness scale is 
nontrivial, and to show this
we will step through each contribution to the beta functions in
the high energy theory (well above the compositeness scale).

The one-loop beta functions for the SM gauge couplings are\footnote{We
use the $SU(5)$ GUT normalization $b_1 = (3/5) b_Y$.}
\begin{equation}
\frac{d}{d t} g_a = \left( b_a^{\rm MSSM} + \Delta b_a \right) g_a^3 \; ,
\end{equation}
where $b_a^{\rm MSSM}$ are the MSSM contributions and the $\Delta b_a$
characterize differences between our model and the MSSM. Above the
compositeness scale our model has no fundamental Higgs
fields. However, the $SU(2)_H$ doublets $T^1,\ldots,T^4$ give exactly
the same contribution to the beta functions as the two Higgs doublets
of the MSSM. Thus the selection of $SU(2)_H$ as the strong gauge group
has the interesting side-effect that the fundamental and composite
states give precisely the same contribution to the SM gauge beta
functions. Since $T^5,\ldots,T^8$ are neutral under the SM, they do not
contribute to the running of the SM gauge couplings.

Our model has two new sectors that contribute to $\Delta b_a$: (1) the
$P$ and $Q$ fields that marry off spectator composite fields, and (2)
the extra doublets $\varphi_{u,d},\overline{\varphi}_{u,d}$ needed to
generate fermion masses and mixings.

The first contribution consists of the $P$ and $Q$ fields, which yields 
\begin{equation}
\Delta b_1 = \frac{3}{5} \ , \quad \Delta b_2 = 1 \ , \quad \Delta b_3 = 0 
\label{b3-eq}
\end{equation}
which corresponds to two $SU(2)_L$ doublets and four
fields with hypercharge $\pm 1/2$.

The second contribution, from the extra doublets $\varphi_{u,d},
\overline{\varphi}_{u,d}$ needed for fermion masses, is
\begin{equation}
\Delta b_1 = \frac{6}{5} \ , \quad \Delta b_2 = 2 \ , \quad \Delta b_3 = 0 
\label{b4-eq}
\end{equation}
which corresponds to four $SU(2)_L$ doublets with hypercharge
$\pm 1/2$.

The total of (\ref{b3-eq}) and (\ref{b4-eq}) is
\begin{equation}
\sum \Delta b_1 = \frac{9}{5} \ , \quad \sum \Delta b_2 = 3 \ , \quad 
\sum \Delta b_3 = 0 \; .
\end{equation}

Coupling unification requires us to add additional matter at the
$\Lambda_H \sim m' \sim M_f$ scale. For instance, we can add three
vector-like pairs of chiral multiplets $D_i (\,\overline{\!D}{}_i)$,
$(i=1,2,3)$, with the quantum numbers of the right-handed down quarks,
{\it i.e.}\/ triplets under SU(3) with $U(1)_Y$ quantum numbers $\pm
1/3$. Then
\begin{equation}
\sum \Delta b_1 = \sum \Delta b_2 = \sum \Delta b_3 = 3 \; .
\end{equation}
This is the same result obtained for a gauge mediation model
with three sets of $\mathbf{5} + \overline{\mathbf{5}}$ messengers.

Like gauge mediation, these extra fields have the appearance of
``completing'' the would-be incomplete SU(5) matter representations.
For example, a gauge mediation model with only messenger quark
doublets $Q_m + \overline{Q}_m$ is sufficient to communicate
supersymmetry breaking, but as these fields do not form a complete
SU(5) representation, additional messenger fields filling up the ${\bf
10}_m$ and $\overline{\bf 10}_m$ must be added to preserve gauge
coupling unification. Unlike gauge mediation, however, the extra color
triplets $D (\,\overline{\!D}{})$ cannot be in the same GUT
representation as $\varphi_{u,d},\overline{\varphi}_{u,d}$, otherwise
dimension-6 triplet-induced proton decay will be too fast. In any
case, we have shown that adding three pairs of color triplets to our
model does not affect the dynamics and yet provides an existence proof
that gauge coupling unification can work just as well as in the
MSSM\@.

It is also important to emphasize that we do not expect large
threshold corrections from passing through the strong
coupling/superconformal sector. Due to holomorphy, the low-energy
gauge couplings are determined only by the bare mass of the heavy
particles that are integrated out~\cite{Arkani-Hamed:1997ut}. This can
also be seen by noting that in supersymmetric theories both the exact
NSVZ beta function \cite{NSVZ,Arkani-Hamed:1997mj} and the decoupling
mass depend on the wave-function renormalization factor, which drops
out from the final result. Therefore, gauge coupling unification is
unaffected even in the presence of strong $SU(2)_H$ dynamics in which
the standard model gauge groups $SU(3)_c \times SU(2)_L \times U(1)_Y$
are perturbatively coupled. The dominant effect is therefore the
threshold correction resulting from potential differences between the
mass of the color triplets, the mass $m_{\rm spect}$ of the
spectators, and the mass $M_f$ of the extra doublets
$\varphi$. Suppose that the same flavor symmetry that ensures the $T^7
T^8$ doublets acquire the mass $m'$ could also be used to determine
the color triplet masses. In this case the threshold corrections are
no larger than $\log m'/M_f$ or $\log m'/m_{\rm spect}$, of the same
order of magnitude as the MSSM or GUT threshold corrections, which is
much smaller than the leading $\log M_{\rm unif}/M_Z$ in the MSSM.

We have shown that gauge coupling unification can be preserved with a
small number of additional matter fields, but it is obvious that we
cannot embed the matter content into a single four-dimensional GUT
group. Unification of the gauge couplings therefore could be due to
string unification or orbifold GUT unification in five
\cite{orbifoldGUT} (or four \cite{ssgut}) dimensions, where the matter
content does not need to fall into a GUT representation
\cite{splitmultiplet}.

\section{Discussion}
\label{sec:discussions}

We have constructed a supersymmetric composite Higgs theory that
solves the supersymmetric little hierarchy problem.  Electroweak
symmetry is broken dynamically through a new gauge interaction that
gets strong at an intermediate scale.  The composite Higgs fields have
a dynamically generated superpotential that has a form similar to the
NMSSM, and hence solves the $\mu$-problem, but with no restriction on
the coupling $\lambda$.  This allows the tree-level Higgs mass to be
much higher, $200$-$450$~GeV, solving the supersymmetric little
hierarchy problem.  The usual lore about upper bounds on the lightest Higgs
boson mass in supersymmetric theories is therefore obviously violated.
With hindsight we see that requiring perturbativity of the Higgs
sector was simply too restrictive. To the best of our knowledge, the
Fat Higgs model provides the first explicit example where the Higgs
sector is composite and yet the dynamics are fully calculable and UV
complete.

There are several interesting future avenues of research.  We used the
Giudice-Masiero mechanism to determine certain mass scales, and
therefore supergravity-mediation was implicit.  For generic choices of
the supergravity-mediated contributions we have the regular
supersymmetric FCNC problem.  One solution is a flavor symmetry,
\emph{e.g.} $U(3)^5$ in \cite{Hall:1990ac}.  Another possibility is to
implement one of several flavor-blind supersymmetry breaking
mechanisms such as gauge mediation \cite{Dine:1995ag}, anomaly
mediation \cite{AMSB} (supplemented by $U(1)$ $D$-terms to make it
viable with UV insensitivity \cite{Dterm}) or its 4D realization
\cite{AMSB4D}, or gaugino mediation in five \cite{gaugino} or four
\cite{gaugino4D} dimensions. It remains to be seen whether these
mediation mechanisms achieve an acceptable mass spectrum and
electroweak symmetry breaking.  These methods of supersymmetry
breaking would also require a different mechanism to naturally
determine the scales. It would also be interesting to explore
unification further in this model, such as whether $SU(2)_{H}$ can be
unified with the other SM gauge groups.

Finally, we have shown that the Higgs mass spectrum is quite unusual.
It is important to study specifically how our Fat Higgs model can be
distinguished from more conventional supersymmetric models at future
collider experiments. Clearly more work is needed. We cannot
overemphasize the importance of next generation experiments being able
to analyze their data with as few theoretical assumptions as possible.

\acknowledgments{We thank Jens Erler, Markus Luty, Michael Peskin and
Aaron Pierce for discussions. RH and DTL thank the Institute for
Advanced Study for hospitality. GDK is a Frank and Peggy Taplin Member
and thanks them for their generous support of the School of Natural
Sciences. This work was supported by the Institute for Advanced Study,
funds for Natural Sciences, as well as in part by the DOE under
contracts DE-FG02-90ER40542 and DE-AC03-76SF00098 and in part by NSF
grant PHY-0098840.}

\appendix

\section{Higgs Sector Contribution to $S$ and $T$}
\label{sec:higgs}

Here we provide expressions for the perturbative contribution of the
Higgs sector to $S$ and $T$. The Higgs sector consists of mass
eigenstates $H^\pm$, $H^0$, $h^0$, and $A^0$ which fit into an
$SU(2)_L$ doublet,
\begin{equation}
\label{doublet}
\left(
  \begin{array}{c}
     H^+ \\
     \frac{1}{\sqrt{2}}\left( \widetilde{H}^0+iA^0 \right)
  \end{array} \right),
\end{equation}
where $\tilde{H}^0 = \cos(\beta-\alpha) h^0 - \sin(\beta-\alpha) H^0$,
and a Standard Model-like neutral scalar $\tilde{h}^0$, which is
orthogonal to $\widetilde{H}^0$. The Higgsinos get a vector-like mass
of order $m_{\rm SUSY}$, so their contribution to $S$ and $T$ is
negligible. $\tilde{h}^0$ will contribute much like a heavy Standard
Model Higgs. However, since it is not a pure mass eigenstate, the
exact contribution is fairly complicated. We approximate this
contribution by taking the weighted sum of the two-loop contributions
extracted from the electroweak observables discussed in
Appendix~\ref{sec:ST}.

Below we summarize the one-loop contribution of the scalar Higgs
doublet. Note that the various components of the doublet all have
different masses, and the neutral scalar $\widetilde{H}^0$ is itself a
linear combination of two mass eigenstates. Taking this mixing into
account, we find:
\begin{eqnarray}
\Delta S &=& \sin^2(\beta-\alpha) F(m_{H^\pm}, m_{H^0}, m_{A^0})
\nonumber \\
&&  + \cos^2(\beta-\alpha) F(m_{H^\pm}, m_{h^0}, m_{A^0}).
\end{eqnarray}
where the function $F$ is defined by
\begin{eqnarray}
  \lefteqn{
    F(m_1, m_2, m_3) } \nonumber \\
  &=& \frac{1}{2\pi} \int_0^1 \!dx\, x(1-x) \log
    \frac{(1-x)m_2^2 + x m_3^2}{m_1^2}.
\end{eqnarray}
Similarly for $T$, we find
\begin{eqnarray}
\Delta T &=& \frac{1}{16\pi m_W^2 s_W^2} \left( \sin^2(\beta-\alpha)
G(m_{H^\pm}, m_{H^0}, m_{A^0}) \right. \nonumber \\
&& \left. + \cos^2(\beta-\alpha) G(m_{H^\pm}, m_{h^0}, m_{A^0}) \right).
\end{eqnarray}
where the function $G$ is defined as
\begin{eqnarray}
  \lefteqn{
    G(m_1, m_2, m_3)}  \nonumber \\
  &&= m_2^2 I(m_3,m_2,m_1,m_2) + m_3^2 I(m_2,m_3,m_1,m_3) \nonumber \\
  &&- m_1^2 I(m_2, m_1,m_1,m_1) - m_1^2 I(m_3,m_1,m_1,m_1) \nonumber \\
\end{eqnarray}
in terms of the integral $I$,
\begin{equation}
I(m_1, m_2,m_3, m_4) = 2\int_0^1\!dx\,x\log \frac{(1-x)m_1^2+x
m_2^2}{(1-x)m_3^2+x m_4^2}.
\end{equation}

\section{$S$-$T$ Contours}
\label{sec:ST}

The experimental constraints on the $S$-$T$ plane can be easily
computed approximately using the following method. We follow the path
of Marciano \cite{Marciano} and Perelstein--Peskin--Pierce \cite{PPP}
to focus on only three observables, $m_W$, $\Gamma_l$, and $s_\ast^2
\approx \sin^2\theta_{\it eff}^{\it lept}$ from the asymmetries as
they are the most accurately measured and sensitive observables to the
oblique corrections.

Expressions for these observables, including their approximate $m_t$,
$\alpha$, and $m_H$ dependence, have been computed by Degrassi and
Gambino~\cite{Degrassi:1999jd}. We add to those expressions the
dependence on $S$ and $T$ as found in Appendix B of Peskin--Takeuchi
\cite{PT}. The LEP Electroweak Working Group recommends $\Delta
\alpha_{\rm had}^{(5)}(m_Z^2) = 0.02761(36)$, including the BES data
as discussed in section 16.3 of \cite{LEPEWWG}, which implies
$\alpha^{-1}(m_Z) = 128.945\pm 0.049$. Expanding to linear order in
$\Delta m_t$ and $\Delta \alpha^{-1}$ about $m_t=174.3$ GeV and
$\alpha^{-1}=128.945$ leads to the expressions
\begin{eqnarray}
    m_W &=& 80.380 + 0.13 \Delta\alpha^{-1} + 0.0061 \Delta m_t - 0.29
    S
  \nonumber \\
  & & + 0.44 T + 0.34 U - 0.058 l_h -0.008  l_h^2, \nonumber \\
    s_\ast^2 &=& 0.23140 - 0.0026 \Delta\alpha^{-1} - 0.000032 \Delta m_t
  \nonumber \\
  & &+  0.0036 S - 0.0025 T + 0.00052 l_h, \nonumber \\
    \Gamma_l &=& 84.011 + 0.12 \Delta\alpha^{-1} + 0.009 \Delta m_t 
  \nonumber \\
  & & - 0.19 S + 0.78 T - 0.054 l_h -0.021 l_h^2, 
\end{eqnarray}
where $l_H = \log(m_H/100~{\rm GeV})$.

For the experimental values, we use \cite{PPP}
\begin{eqnarray}
  m_W &=& (80.425 \pm 0.034)~{\rm GeV},\\
  s_\ast^2 &=& 0.23150 \pm 0.00016, \\
  \Gamma_l &=& (83.984 \pm 0.086)~{\rm MeV}.
\end{eqnarray}
Then $\chi^2$ is defined as
\begin{eqnarray}
\lefteqn{    \chi^2 = \frac{(m_W - 80.425)^2}{0.034^2} +
  \frac{(s^2-0.23150)^2}{0.00016^2} }\nonumber \\ \nonumber \\
  & &+ \frac{(\Gamma_l -
    83.984)^2}{0.086^2} + \frac{(\Delta \alpha^{-1})^2}{0.049^2} +
  \frac{(\Delta m_t)^2}{5.1^2}, 
\end{eqnarray}
which is first minimized with respect to $\Delta\alpha^{-1}$ and
$\Delta m_t$ for each $(S,T)$. This expression for $\chi^2$ yields
contours that agree very well with those by the Particle Data Group
\cite{Hagiwara:fs}.


\end{document}